\pdfoutput=1
\documentclass[pr,twocolumn]{revtex4}
\usepackage{amssymb}
\usepackage{amsmath}
\usepackage{mathptmx}
\usepackage{color}
\usepackage{euscript}
\usepackage{eufrak}
\usepackage{hyperref}
\usepackage{graphicx}
\usepackage{bm}

\begin{document}

\title{Polarization conversion spectroscopy of hybrid modes}

\author{A. Yu. Nikitin,$^{1,2,*}$ David Artigas,$^{3,4}$ Lluis Torner,$^{3,4}$ F.~J.~Garc\'{i}a-Vidal,$^5$ and L. Mart\'{\i}n-Moreno$^{1}$}

\address{
$^1$Instituto de Ciencia de Materiales de Arag\'{o}n and Departamento de F\'{i}sica de la Materia Condensada,\\
CSIC-Universidad de Zaragoza, E-50009, Zaragoza, Spain
\\
$^2$A. Ya. Usikov Institute for Radiophysics and Electronics, Ukrainian Academy of Sciences, 61085 Kharkov, Ukraine
\\
$^3$ICFO-Institut de Ciencies Fotoniques, Mediterranean Technology Park, 08860 Castelldefels (Barcelona), Spain
\\
$^4$Department of Signal Theory and Communications, Universitat Polit\'{e}cnica de Catalunya, 08034 Barcelona, Spain
\\
$^5$ Departamento de F\'{i}sica Te\'{o}rica de la Materia Condensada, Universidad Aut\'{o}noma de Madrid, E-28049 Madrid, Spain
\\
$^*$Corresponding author: alexeynik@rambler.ru }

\begin{abstract}
Enhanced polarization conversion in reflection for the Otto and
Kretschmann configurations is introduced as a new method for
hybrid-mode spectroscopy. Polarization conversion in reflection
appears when hybrid-modes are excited in a guiding structure
composed of at least one anisotropic media. In contrast to a dark
dip, in this case modes are associated to a peak in the converted
reflectance spectrum, increasing the detection sensitivity and
avoiding confusion with reflection dips associated with other
processes as can be transmission.
\end{abstract}

\maketitle

\begin{figure}[ht]
\centerline{
\includegraphics[width=8.3cm]{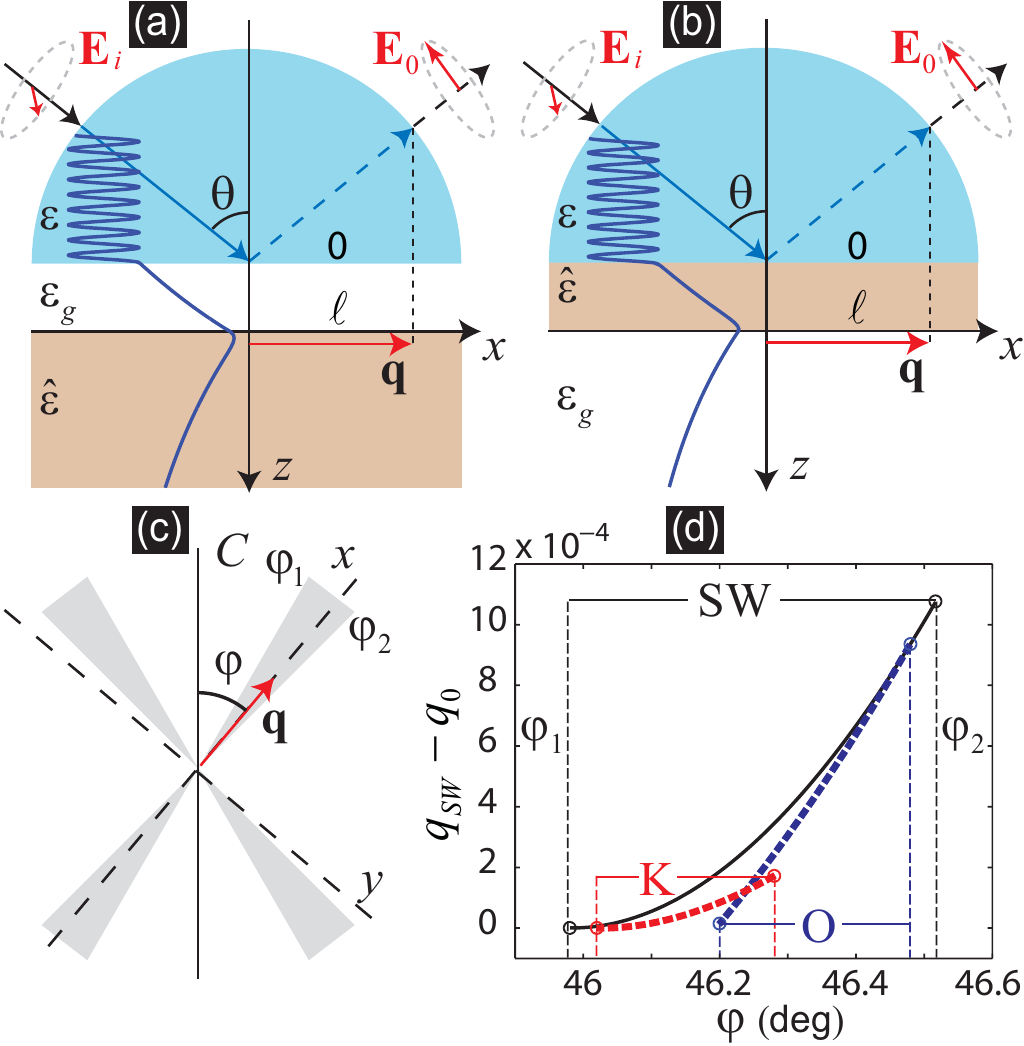}}
\caption{(Color online) Geometry of the studied systems corresponding to the (a) O and (b) K configuration. (c) dashed sectors schematically show the values of angle $\varphi$ at which Dyakonov SWs exist. (d) effective index $q_{SW}$ in the range of SW existence for the original Dyakonov (black solid line), O (blue dashed line) and K (red dashed line) configurations for the following parameters: $\epsilon_g = 4.41$ (Ta$_2$O$_5$), $\epsilon = 6.7$ (ZnSe), $\epsilon_\perp = 3.97$, $\epsilon_\parallel = 4.9$ (YVO$_4$), $\ell=4\lambda$, $q_0=\sqrt{\varepsilon_g}$.}
\end{figure}

Waves incident onto the surface of an anisotropic medium can get
reflected with a polarization orthogonal to the incident one
\cite{reflection_CM91}. In natural crystals, this conversion of
polarization states is small for incidence below the critical
angle \cite{polariz_CM92} and increases under total reflection
conditions \cite{critang_JMO93}. Total conversion has been
predicted in corrugated structures \cite{sinusoidal_PRB94} and in
metallic interfaces supporting plasmons \cite{Bryan-Brown90,
Kats05, plasmon09}. Structures using anisotropic thin films have
shown an enhancement of this polarization conversion
\cite{thinfilm_OC06}. Importantly, similar structures can support
different kinds of hybrid guided modes. A special case corresponds
to surface waves (SWs) supported at the interfaces between
anisotropic and isotropic media \cite{D'yakonov88}. These SWs,
referred to as Dyakonov SWs, are hybrid waves existing under
special conditions, which have been recently observed thanks to
the presence of the polarization conversion effect in the
Otto-Kretschmann excitation scheme \cite{Takayama09}. Under this
scheme, the usual dip in a bright background observed in
reflection (conserving polarization) was substituted by a
bright-peak in a dark background for the orthogonal polarization.
Additionally, reflection dips which are not associated to modes of
the structure, as can be transmission, did not result in a peak in
the polarization converted image. This provided a higher contrast
and specificity, making possible the observation of Dyakonov SWs.

In addition to increasing contrast, the results in Ref.\cite{
plasmon09, thinfilm_OC06, Takayama09} suggest that, in the
Otto-Kretschmann configuration, the hybrid nature of any existing
mode is related to the enhanced polarization conversion effect.
The aim of this paper is to theoretically demonstrate that the
mode excitation is related to polarization conversion, which can
be used as a new method for hybrid mode spectroscopy.

\begin{figure*}[htb]
\centerline{
\includegraphics[width=16.6cm]{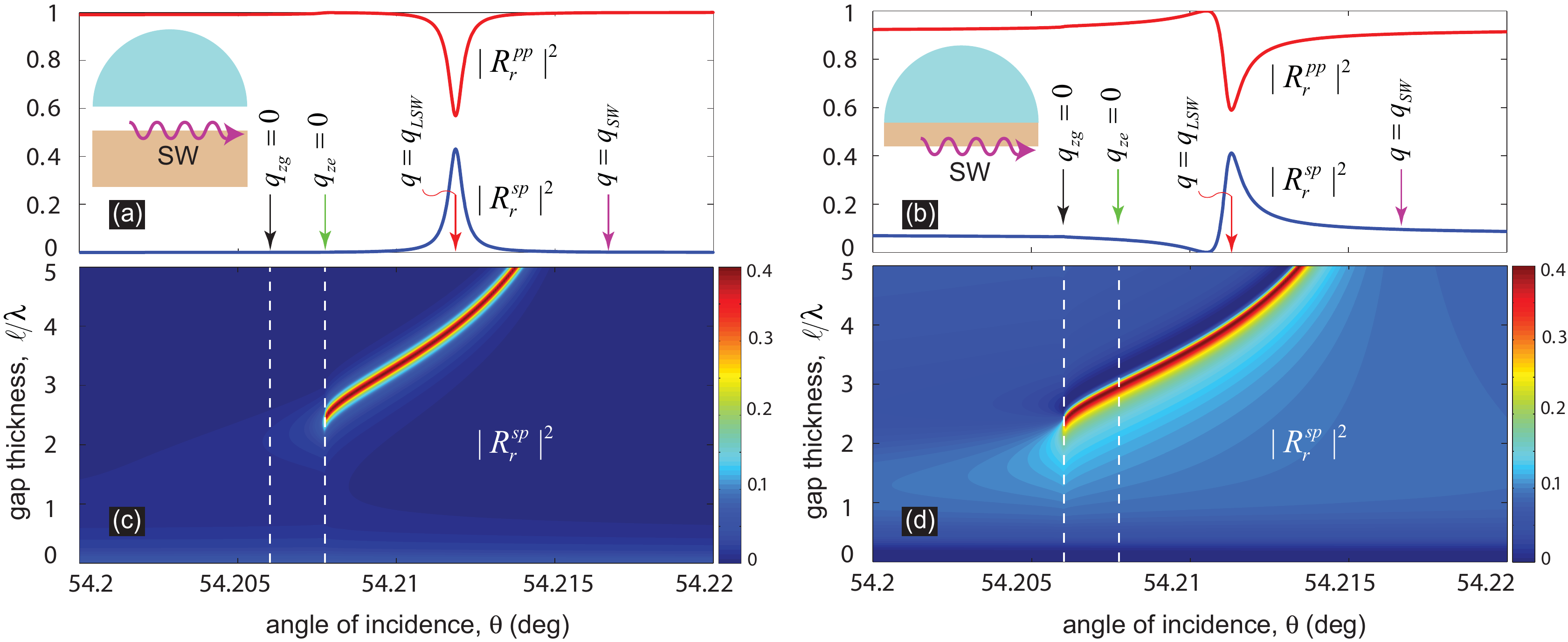}}
\caption{(Color online)(a,b) Square modulus of the polarization
RCs, as functions of the incidence angle $\theta$, for a gap
thickness $\ell = 4\lambda$ and for  $\varphi = 46.25^\circ$. (a)
is for O configuration, while (b) is for K configuration. (c,d)
render the contour plot for the cross-polarization RCs squared
modulus as a function of both $\theta$ and $\ell$. All the parameters
are the same as in Fig.~1.}
\end{figure*}

Consider a plane arbitrarily-polarized monochromatic wave with the
wavevector $\mathbf{k}_{inc}$ incident from a prism with
dielectric permittivity $\epsilon$ onto a uniaxial crystal
characterized by a dielectric tensor $\hat{\epsilon}$ with
longitudinal, $\epsilon_\|$, and orthogonal, $\epsilon_\perp$,
components. The Otto (O) geometry is achieved when the uniaxial
crystal is separated from the prism by a dielectric medium (gap),
with width $\ell$ and permittivity $\epsilon_g$ [Fig.~1(a)]. The
Kretschmann (K) geometry is obtained when the crystal with
thickness $\ell$ is sandwiched between the prism and the
dielectric medium [Fig.~1(b)]. The laboratory axes are chosen so
that the axis $\bf{z}$ is orthogonal to the media interfaces and
the axis $\bf{x}$ is oriented an angle $\varphi$ relatively to the
optic axis $C$ of the crystal. The interface of the prism
coincides with the $z=0$ plane. The electric field in the prism is
written in a compact form as

\begin{equation}\label{fields}
\mathbf{E}_{m}(\textbf{r}) = \sum_{\sigma}\mathbf{e}_i^\sigma
E_i^{\sigma}e^{i\mathbf{k}_{inc}\mathbf{r}} + \sum_{\sigma,\sigma'}\mathbf{e}_r^\sigma
R_r^{\sigma\sigma'}E_i^{\sigma'}e^{i\mathbf{k}_r\mathbf{r}}.
\end{equation}
Here $\sigma=s(p)$ specifies the polarization. The index ``$i$''
is related to the incident wave. $E_i^{\sigma}$ are the
polarization amplitudes of the incident wave and ``$r$'' stays for
the reflected one; $R_r^{\sigma\sigma'}$ are the reflection
coefficients (RCs). The polarization vectors are $\mathbf{e}^s =
\mathbf{e}^y$ and $\mathbf{e}^p = \mathbf{e}^s\times\mathbf{q}/q$,
where $\mathbf{q} = \mathbf{k}_t/g$ is the dimensionless
wavevector component parallel to the interface, and $g =
\omega/c$. In the laboratory coordinate system
$q=\sqrt{\epsilon}\sin\theta$, $\theta$ being the incident angle.
The fields inside the isotropic gap (in the case of the O
geometry) and inside the substrate (for the K geometry) are
decomposed using the same polarization basis. Inside the crystal
the unit vectors for the extraordinary and ordinary waves are
$\textbf{e}^e =
\varepsilon_\bot\mathbf{e}^x-(\mathbf{e}^x\cdot\mathbf{q}_e)\mathbf{q}_e$
and $\textbf{e}^o = \mathbf{e}^C\times\mathbf{q}^o$ respectively.
The $z$-components of the dimensionless wavevectors $q_{z\alpha} =
k_{z\alpha}/g$, where $\alpha$ specifies the medium,  are
$q_{zi}=-q_{zr}=\sqrt{\epsilon}\cos\theta$,
$q_{zg}=\pm\sqrt{\epsilon_g-q^2}$,
$q_{zo}=\pm\sqrt{\epsilon_\perp-q^2}$,
$q_{ze}=\pm\sqrt{\epsilon_\parallel-q^2[\sin^2\varphi+(\epsilon_\parallel/\epsilon_\perp)\cos^2\varphi]}$.
The sign choice depends upon the propagation direction of the
corresponding wave.

Matching the tangential components of the fields at the boundaries
$z=0$ and $z=\ell$, we arrive at the system of linear equations
for the unknown RCs. Then the eigen modes of the system are the
solutions of the homogeneous system of equations, when
$E_i^\sigma=0$. The modes dispersion relation is obtained from the
zeros of the determinant of the system. These equations reduce to
the solution for the surface waves studied by Dyakonov
\cite{D'yakonov88} in the limit $\ell \rightarrow \infty$. Recall
that these hybrid SWs (with purely imaginary $q_{zg}$, $q_{zo}$,
and $q_{ze}$ for the given $q$) exist in a certain range of angles
$\varphi$ under the condition
$\epsilon_\parallel>\epsilon_g>\epsilon_\perp$ [Fig.~1(c,d)]. An
incident wave from one of the half-spaces cannot couple to the SW
directly since the phase velocity of the SW is less than that of
the incident wave. The O and K configurations solve the problem by
adding a prism with $\epsilon>\epsilon_g, \epsilon_\parallel,
\epsilon_\perp$. Now, the radiation leakage is added to the SW and
the wave incident from the prism at angles exceeding the total
internal reflection one, $\cos\theta>\sqrt{\epsilon_g/\epsilon}$,
can couple to the SW [Fig.~1(a,b)]. However, the radiation leakage
restricts the angular interval of the SW existence. For the O
configuration the dispersion curve is mainly cut from the lower
angles, while for the K geometry the curve is cut from the higher
angles.

Returning to the inhomogeneous case, we would like to emphasize
that, as follows directly from the system of equations,
$|R_r^{sp}| = |R_r^{ps}|$ and $|R_r^{ss}| = |R_r^{pp}|$ when the
fields are evanescent both in the isotropic media and in the
crystal. Thus, the reflection of the plane wave is symmetric
relative to the polarizations of the incident and reflected waves.
An example of the SW excitation for a YVO$_4$ crystal is shown in
Fig.~2. For this crystal, and using Ta$_2$O$_5$ as the isotropic
medium, the pure Dyakonov SW exists in the angular range
$\varphi\in [45.98^\circ,46,52^\circ]$. The vertical arrows both
in (a) and (b) indicate the positions of the angle for the pure
Dyakonov SW ($q = q_{SW}$), leaky Dyakonov SW ($q = q_{LSW}$) and the branch points for both the extraordinary wave,
$q_{ze}=0$, and the isotropic medium, $q_{zg}=0$. The branch point
of the ordinary wave, $q_{zo}=0$, is far below the interval of
incident angles $\theta$ shown in the figure. The spectral
position of the leaky Dyakonov SW virtually coincides with the
angle of RCs maximum, where enhanced polarization conversion takes
place [Fig.~2 (a,b)]. The width of the resonance curve depends
upon the gap thickness as shown in Fig.~2 (c,d). For each fixed
$\theta$ there is an optimal value of $\ell$ corresponding to the
best compromise between coupling strength and radiation leakage.
Interestingly, when $\ell$ increases, the required angle of
incidence evolves towards the value corresponding to the pure
Dyakonov SW, demonstrating that polarization conversion is related
to the hybrid mode in the structure.

\begin{figure}[htb]
\centerline{
\includegraphics[width=8.3cm]{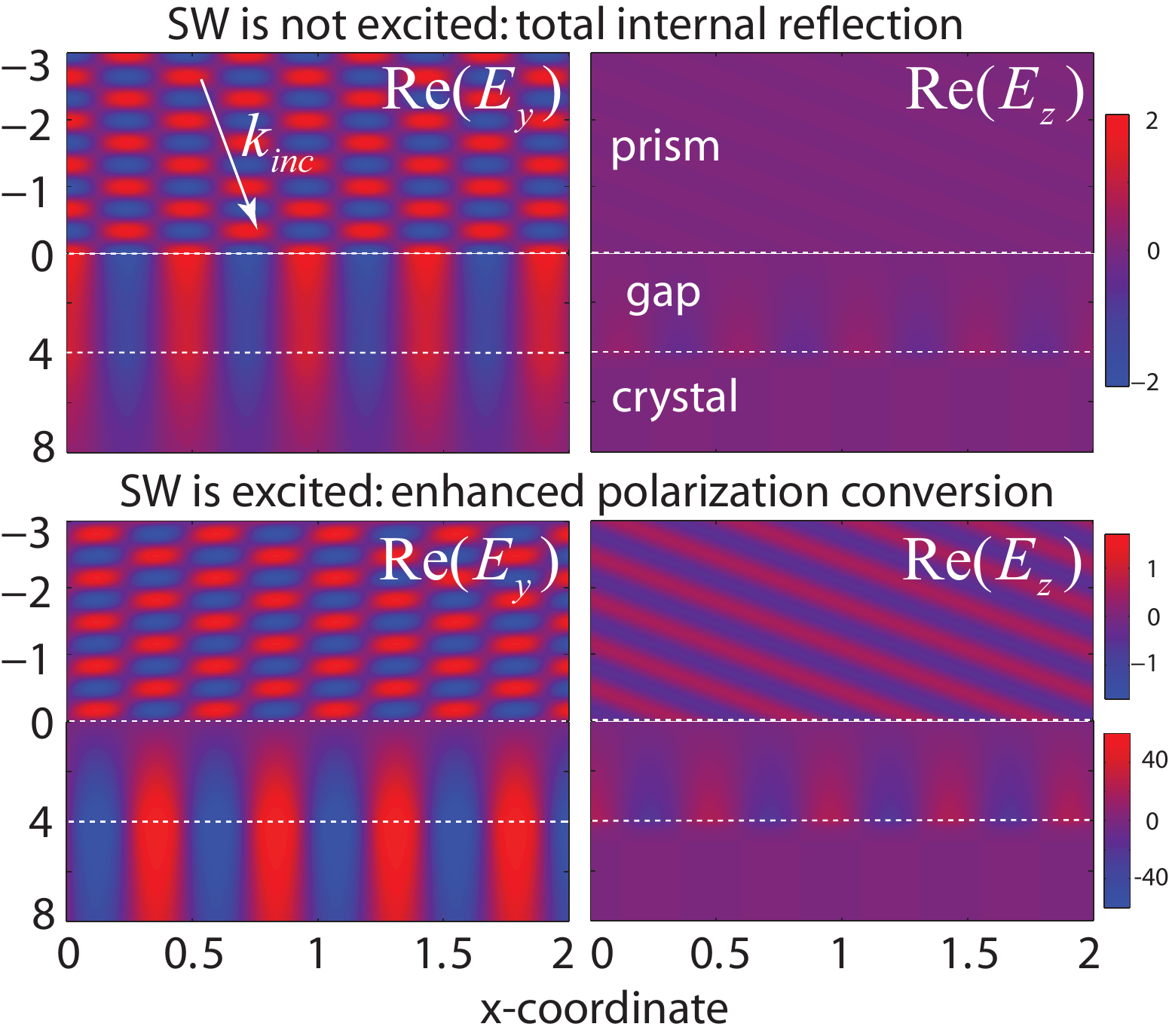}}
\caption{(Color online) The spatial distribution of the real part of the electric field in the O configuration. The calculations for the upper
panels are done at $\theta=54.23^\circ$, while for the lower panel the incidence angle is $\theta=54.212^\circ$. In all panels $\varphi = 46.25^\circ$ and $\epsilon_g$, $\epsilon$, $\hat{\epsilon}$, $\ell$ are the same as in Fig.~1.}
\end{figure}

The transformation process and excitation of the SW can be
visually demonstrated by plotting the instant spatial field
distribution (Fig.~3). In this figure we show two projections of
the electric field when the incident wave is $s$-polarized
(incident field with $y$-component only) in the O configuration.
The two upper insets show the nonresonant case, when the
SW is not excited. Here a strong internal reflection takes place,
resulting in an interference pattern for the $E_y$ component. In
this case, which does not show polarization conversion, all the
fields below the prism are evanescent and inside the crystal they
are distributed between the ordinary and extraordinary components.
Under the resonant conditions, shown in two lower insets, the
increase of the field amplitude at the boundary between the
crystal and the gap demonstrates the excitation of the SW. This is
connected with a reduction in the $y$-component of the reflected
wave (resulting in a lower contrast interference pattern) that is
converted into the orthogonal polarization, leading to a strong
reflection in $E_z$ component.

\begin{figure}[htb]
\centerline{
\includegraphics[width=8.3cm]{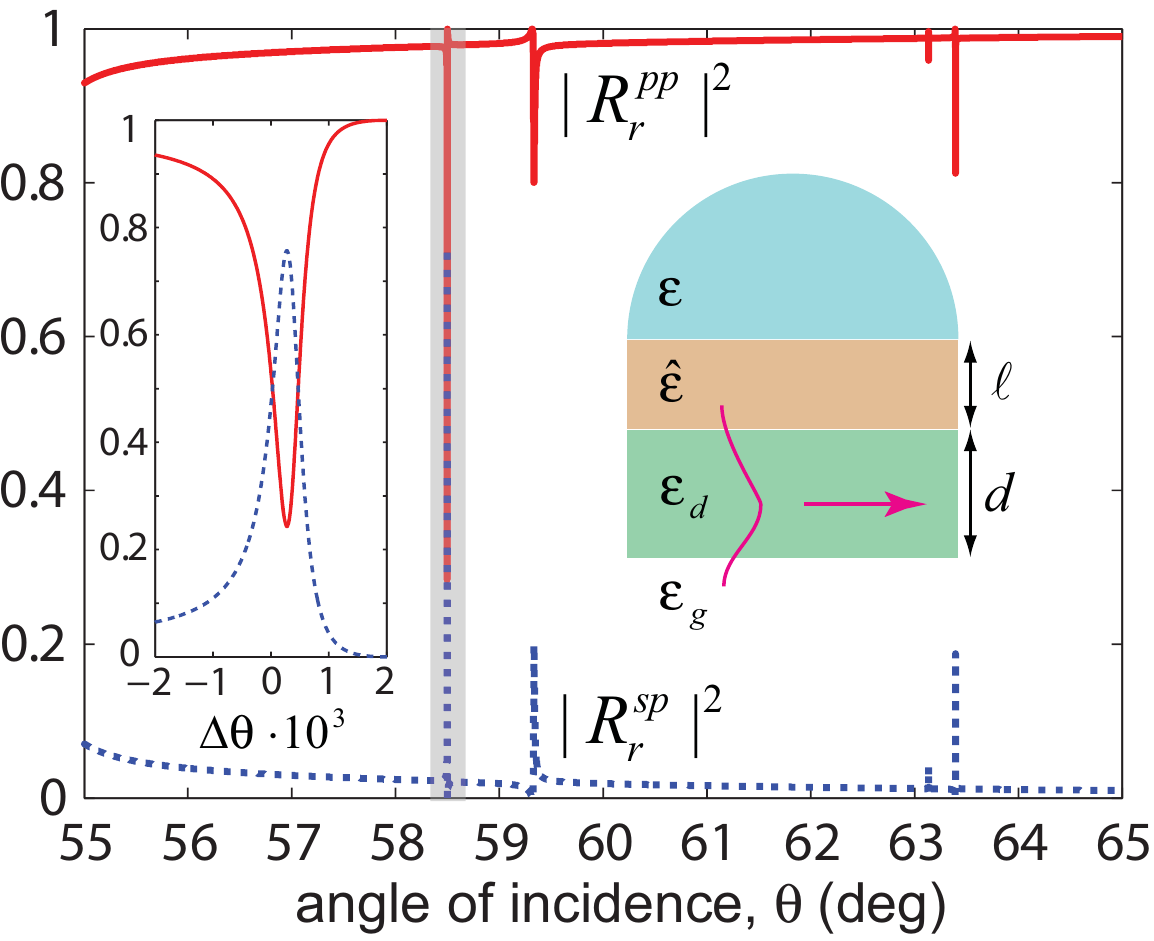}}
\caption{(Color online) Reflection coefficients for hybrid mode spectroscopy in a waveguide in K geometry with $\ell = 0.7\lambda$, $d =\lambda$, $\epsilon_d = 5.5$; and $\epsilon_g$, $\epsilon$, $\hat{\epsilon}$ as in Fig. 1. The inset represents the zoom of the highlighted region with $\Delta \theta = \theta - \theta_0$ with $\theta_0 = 58.497^\circ $.}
\end{figure}

The central point of this Letter is that polarization conversion
spectroscopy is not only restricted to Dyakonov surface waves, but
to other structures supporting hybrid modes. For example, Fig.~4
shows polarization conversion for the simpler case of a waveguide,
where four peaks corresponding to the two first transversal electric (TE)- and
transversal magnetic (TM)-dominant hybrid modes are clearly shown. Polarization
conversion has also been experimentally observed in a more complex
situation involving metals \cite{plasmon09}, where the excited plasmon was a hybrid SW which was mainly a TM mode\cite{mihalache94}.

These results demonstrate the link between excitation of hybrid
modes and enhanced polarization conversion in Otto-Kretschmann
geometries. However, note that analogous resonance effects take
place when hybrid modes are excited not using a prism but by a
periodical structure formed on the surface of the crystal. In that
case the polarization conversion can be observed both in the zero
and higher diffraction orders. As a result, polarization
conversion can be used in all these configurations as a new
hybrid-mode spectroscopic method.

The authors acknowledge financial support from the Spanish
Ministry of Science under projects No. MAT2009-06609-C02, No.
CSD2007-046-Nanolight.es and  Spanish MCyT project FIS2006-10045.
A.Y.N. acknowledges MICINN for Juan de la Cierva Grant.

\end{document}